\documentclass[a4paper, conference]{IEEEtran}
\renewcommand{\baselinestretch}{0.99}
\usepackage{xcolor}
\usepackage{balance}
\usepackage{amsmath,flushend}
\usepackage{subcaption}
\ifCLASSINFOpdf
  \usepackage[pdftex]{graphicx}
\else
  \usepackage[dvips]{graphicx}
\fi
\ifCLASSOPTIONcompsoc
 \usepackage[caption=false,font=normalsize,labelfont=sf,textfont=sf]{subfig}
\else
 \usepackage[caption=false,font=footnotesize]{subfig}
\fi
\def\BibTeX{{\rm B\kern-.05em{\sc i\kern-.025em b}\kern-.08em
    T\kern-.1667em\lower.7ex\hbox{E}\kern-.125emX}}

\begin{document}
\title{Indoor Channel Characterization with\\ Extremely Large Reconfigurable Intelligent Surfaces at $300$ GHz}
\author{\IEEEauthorblockN{
Fabio Cardoso$^{1,2}$, S\'{e}rgio Matos$^{1,2}$, Lu\'{i}s M. Pessoa$^{3,4}$, and George C. Alexandropoulos$^{5,6}$}     \\        
\IEEEauthorblockA{
$^1$Department of Information Science and Technology, University Institute of Lisbon, Portugal\\
$^2$Instituto de Telecomunicações, Portugal\\
$^3$INESC TEC, Portugal, $^4$Faculty of Engineering, University of Porto, Portugal\\
$^5$Department of Informatics and Telecommunications, National and Kapodistrian University of Athens, Greece,\\ 
$^6$Department of Electrical and Computer Engineering, University of Illinois Chicago, IL, USA}\\
\IEEEauthorblockA{e-mails: \{fabio.cardoso, sergio.matos\}@iscte-iul.pt, luis.m.pessoa@inesctec.pt, alexandg@di.uoa.gr}
}

\maketitle

\begin{abstract}
The technology of Reconfigurable Intelligent Surfaces (RISs) is lately being considered as a boosting component for various indoor wireless applications, enabling wave propagation control and coverage extension. However, the incorporation of extremely large RISs, as recently being considered for ultra-high capacity industrial environments at subTHz frequencies, imposes certain challenges for indoor channel characterization. In particular, such RISs contribute additional multipath components and their large sizes with respect to the signal wavelength lead to near-field propagation. To this end, ray tracing approaches become quite cumbersome and need to be rerun for different RIS unit cell designs. In this paper, we present a novel approach for the incorporation of RISs in indoor multipath environments towards their efficient channel characterization. An $100\times100$ RIS design with $2$-bit resolution unit cells realizing a fixed anomalous reflection at 300 GHz is presented, whose radar cross section patterns are obtained via full-wave simulations. It is showcased that the RIS behavior can be conveniently approximated by a three-ray model, which can be efficiently incorporated within available ray tracing tools, and that the far-field approximation is valid for even very small distances from the RIS.
\end{abstract}

\vskip0.5\baselineskip

\section{Introduction}\label{sec:Introduction}
The THz frequency band, in the range of $0.1-10$ THz~\cite{ETSI_THz}, is currently considered as one of the candidate technologies for the upcoming sixth Generation (6G) of wireless networks~\cite{6gvision}, due to its extensive unlicensed bandwidth that can enable ultra high capacity outdoor backhaul links as well as indoor industrial and immersive applications with increased confidentiality, robustness to interference, and reduced latency. However, wireless operations at those frequencies are subject to high penetration loss, which can be critical even for very small link distances. To confront with this challenge, and thus, increase the coverage of THz communications. localization, and sensing~\cite{THs_loc_survey}, extremely large Multiple-Input Multiple-Output (MIMO) systems are being considered~\cite{XLMIMO_tutorial,Shlezinger2021Dynamic,HMIMO_survey}, which are capable of realizing highly directive beamforming. Among the available MIMO solutions belongs the emerging technology of Reconfigurable Intelligent Surfaces (RISs)~\cite{RISoverview2023,EURASIP_RIS}, which offers over-the-air wave propagation control~\cite{alexandg_2021_all,CPH_TWC_2022,SIM_JSAC_2023,RSR2025,OTA2025} to improve and even enable, with cost- and energy-efficient hardware, a wide variety of wireless communications~\cite{Alexandropoulos2022Pervasive}, localization~\cite{Keykhosravi2022infeasible}, and integrated sensing and communications applications~\cite{RIS_ISAC_SPM}. Efficient multi-functional RIS designs and operations schemes for THz wireless applications are recently being developed~\cite{TERRAMETA_Eucap2024,RIS_THz_terrameta,ZML_AWPL_2024,AJG2024,TERRAMETA_website}. 

The characterization and modeling of various high frequency indoor channels with either Line-Of-Sight (LOS) or Non-LOS (NLOS) conditions has been recently receiving substantial research interests, as a means to understand pathloss, delay and angular spread, intra- and inter- cluster characteristics, which can be exploited for efficient signal processing designs\cite{PE_ACCESS_2020, VKJ_WCNC_2022, Model_THz_data_centers, CYY_CST_2022}. A channel model for D-band considering blocking from humans, doors, and partitions for distances up to $10.6$ m was presented in~\cite{PE_ACCESS_2020} that was based on the extended Saleh-Valenzuela model. Large-scale parameters, multipath, directions of arrival and departure with both sides beam steering, and clustering behavior were characterized. A factory hall with two fifth Generation (5G) radio heads in different positions was characterized using both ray tracing and statistical channel models in~\cite{VKJ_WCNC_2022}. A measurement-based hybrid channel model for THz communications among data center use case was introduced in~\cite{Model_THz_data_centers}. In particular, an analytic pathloss approximation was derived that was combined with ray-optical
channel predictions. However, indoor channel characterization at THz frequencies in the presence of realistic extremely large RISs, which contribute dynamically additional multipath components, is only recently receiving research attention~\cite{TERRAMETA_website}.

In this paper, we consider an NLOS indoor communication scenario at $300$ GHz with an $100\times100$ RIS optimized to offer a static non-specular reflection to incoming normal plane waves. We first present the design of our extremely large, with respect to the signal wavelength, RIS comprising $2$-bit resolution unit cells, and then, evaluate its Radar Cross Section (RCS) via full-wave simulations. It is first showcased that the RIS behavior can be conveniently approximated by a three-ray model, which can be efficiently incorporated within available ray tracing tools. In addition, we have found that, although the receive antenna was placed on the near-field region of the RIS, the far-field approximation is quite reasonable for much lower distances, indicating that the proposed approach can be effective for indoor RIS-assisted THz wireless scenarios.

\section{Indoor System Model and Problem Description}\label{sec:Introduction}
\begin{figure}[t!]
	\centering
	\includegraphics[width=0.9\linewidth]{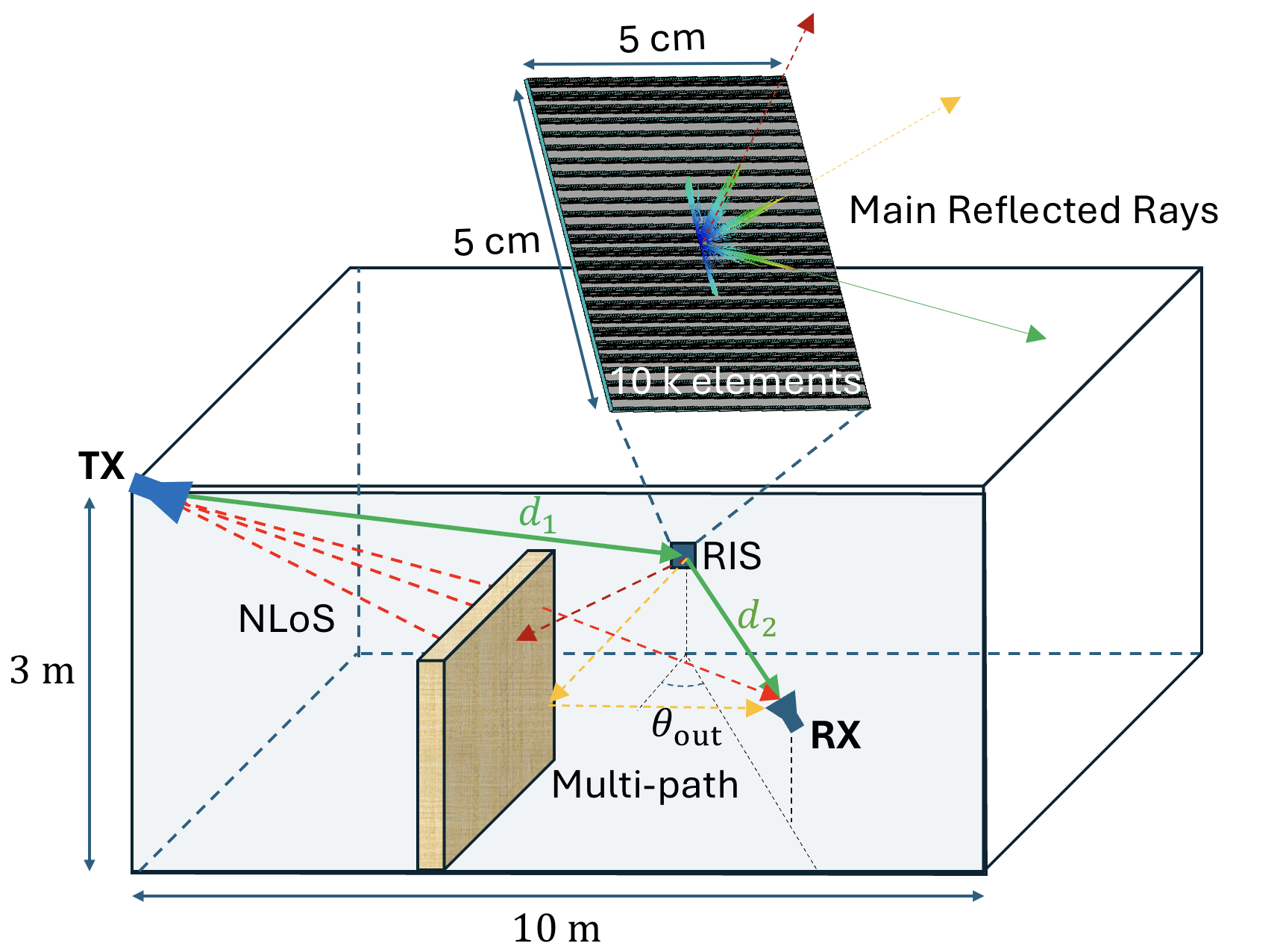}
	\caption{\small{The considered indoor scenario for NLoS wireless communications assisted by an RIS at $304$ GHz. The room dimensions are $10\times10\times3$ m$^3$ and the RIS comprising $100\times100$ unit elements is a $5\times5$ cm$^2$ metasurface designed to facilitate the the TX-RX link.}}
	\label{fig:NLOS_indoor_RIS}
\end{figure}
We consider the NLOS indoor communication scenario of Fig.~\ref{fig:NLOS_indoor_RIS} at sub-THz frequencies, where a single-antenna Transmitter (TX) communicates with a single-antenna Receiver (RX) through an RIS mounted on the wall. The metasurface is of size $5\times5$ cm$^2$ consisting of $100\times100$ sub-wavelength-spaced elements of reconfigurable reflection states. In this paper, we focus on the characterization of an RIS-parameterized NLOS indoor channel for fixed phase configurations of the metasurface. To this end, we have particularly designed the static RIS to redirect incoming normal plane waves into the non-specular reflection $\theta_{\rm out}=30^\circ$ at $304$ GHz. Note that a realistic indoor scenario implies distances of the order of tens of meters, which indicates electrically extremely large RIS aperture for compensating the high propagation pathloss. 

To accurately characterize such indoor propagation channels, a multidomain analysis is required. Multipath characterization using ray tracing algorithms has been lately experimentally proven to be effective at THz frequencies~\cite{Model_THz_data_centers,CYY_CST_2022}. However, the incorporation of RISs into such ray tracing models is still an open research problem. For a realistic description of this problem, a proper characterization (i.e., amplitude and phase) of the rays emerging from the metasurface is required. The typical approach is to consider idealized behavior of the RIS, which, however, may oversimplify channel characterization. On the other hand, full-wave simulations provide a reliable means for the RIS characterization, allowing us to extract equivalent rays from the RCS patterns; see the inset of Fig.~\ref{fig:NLOS_indoor_RIS}. This characterization approach entails the acquisition of all three-dimensional rays emerging from the metasurface, which would heavily overload any ray tracing algorithm if considered in synergy. Fortunately, reasonable approximations can be made to simplify this problem. We will particularly show via full-wave simulations for the designed extremely large RIS, that there exist three main directions where the energy is scattered within our indoor scenario in Fig.~\ref{fig:NLOS_indoor_RIS}: \textit{i}) the RX direction $\theta_{\rm out}$ that corresponds to the RIS main beam (green ray); \textit{ii}) the spurious symmetric beam at the angle $-\theta_{\rm out}$ (red dashed ray); and \textit{iii}) the specular reflection leftover (orange dashed rays). 
 
\section{RIS Design at THz Frequencies}
In this section, we present the design of a static RIS at $304$ GHz that is intended to tilt by $30^\circ$ incoming normal plane waves. A $2$-bit unit cell design was chosen which has been shown to provides a good compromise between number of reconfigurable switch elements and reflection performance~\cite{TERRAMETA_Eucap2024}.

\subsection{Unit Cell Design}
\begin{figure}[t!]
	\centering
	\includegraphics[width=0.9\linewidth]{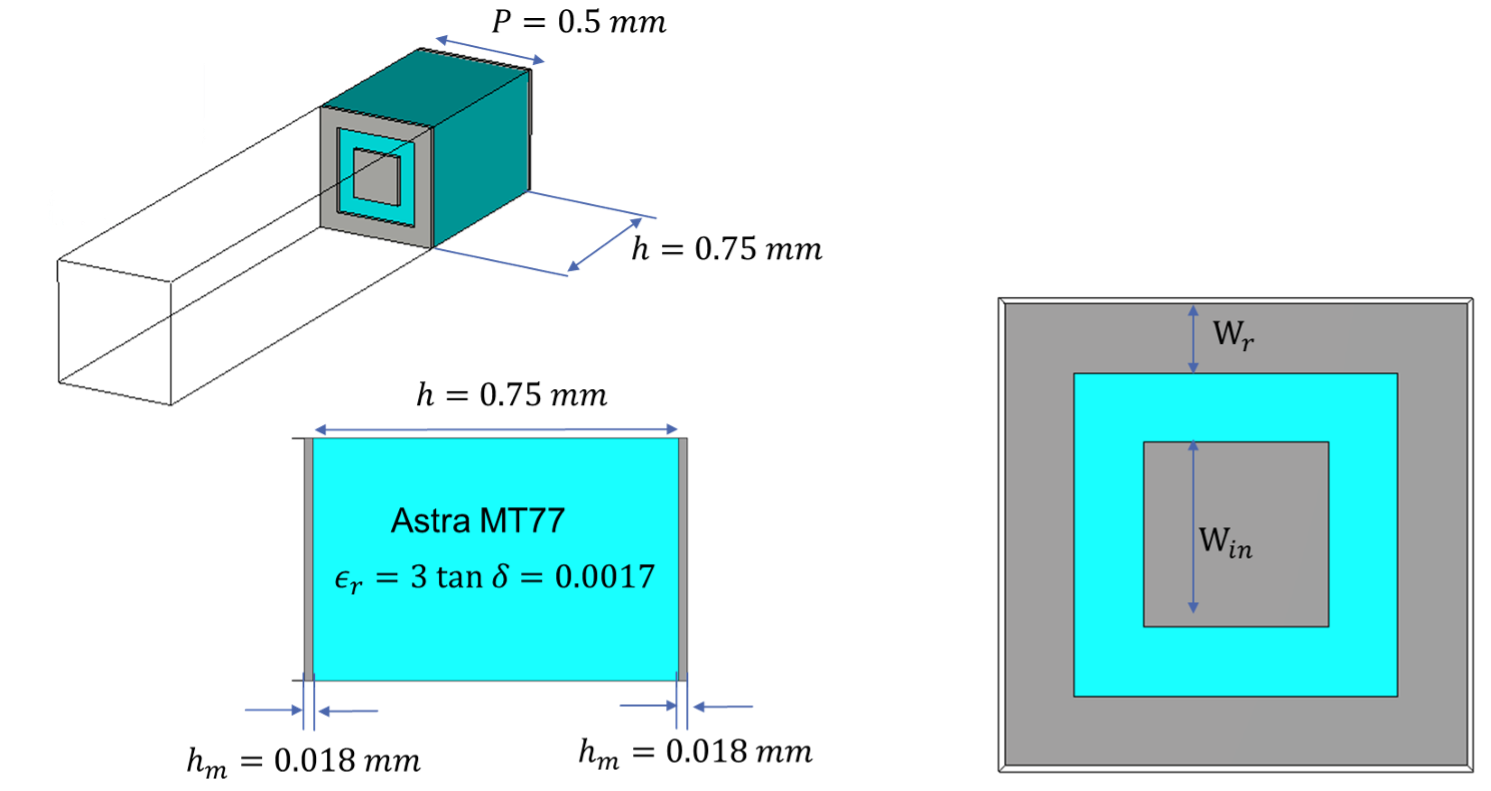}
	\caption{\small{The geometry of the designed single-layer patch-based unit cell at $304$ GHz.}}
	\label{fig:unit_cell}
\end{figure}
\begin{figure}[t!]
	\centering
	\includegraphics[width=0.9\linewidth]{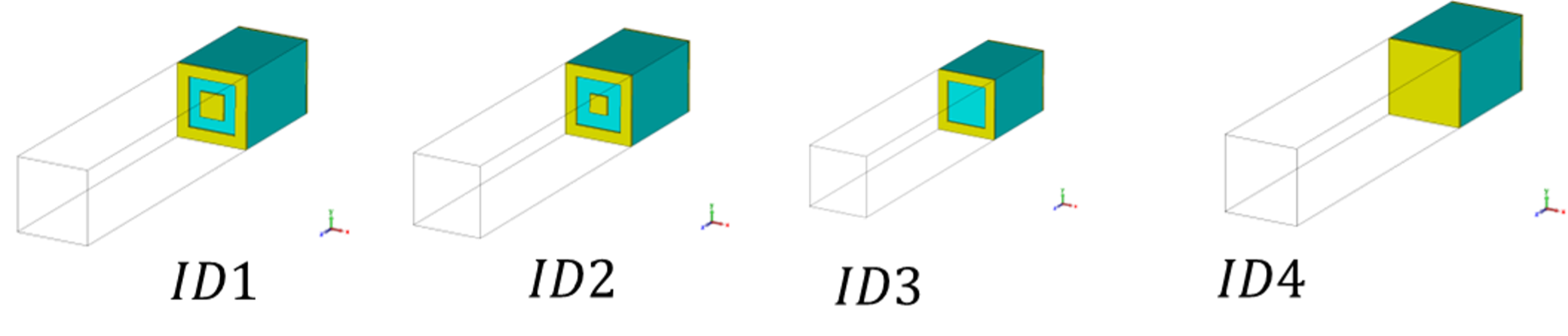}
	\caption{\small{Overview of the unit cell design corresponding to the 4 different states.}}
	\label{fig:unit_cell_states}
\end{figure}

It is quite challenging to produce vias and multilayer structures at THz frequencies, thus, single-layer patch-based unit cells are usually designed, as the one depicted in Fig.~\ref{fig:unit_cell}. The considered substrate is the dielectric Astra MT77 with the standard thickness of $0.75$ mm, which is sufficient to ensure enough mechanical strength for the RIS. The remaining design freedom relies on the patterns of the top layer that provide the required discrete set of phase states. Rectangular shapes are more favourable for production proposes, which led us to fix two building blocks: \textit{i}) squared metallic patch with size $w_{\rm in}$; and \textit{ii}) an outer metallic ring with thickness $w_{\rm r}$. One key point that needs to be accounted for is that, at these high frequencies, small fabrication imperfections can significantly affect the response of the unit cell. We have confirmed that the developed unit cells are stable enough for dimension variations of the order of $20$ $\mu$m; the latter is a reference value for the maximum precision with conventional manufacturing on Printed Circuit Board (PCB). The designed $2$-bit phase resolution unit cell is illustrated in Fig.~\ref{fig:unit_cell_states} for the 4 different states, with the reflection phase given in Table \,\ref{tab:table_states_dimensions}. We also confirm that the reflection amplitude is below $-0.5$ dB. 

\begin{table}[ht!]
\centering
\begin{tabular}{|c||c|c|c|}
    \hline
    ID & $w_{\rm r}$ (mm) & $w_{\rm in}$ (mm)& arg($S_{11})$ (in degrees) \\
    \hline\hline
    1 & 0.08 & 0.17 & 0 \\
    \hline
    2 & 0.08 & 0.13 & 96 \\
    \hline
    3 & 0.08 & 0 & 184 \\
    \hline
    4 & \multicolumn{2}{c}{All metal} & 273 \\
    \hline
\end{tabular}
\caption{\small{Dimensions of the unit cells for the different states and corresponding phase delays.}}
\label{tab:table_states_dimensions}
\end{table}

\subsection{Metasurface Design}
\begin{figure}[t!]
	\centering
	\includegraphics[width=0.9\linewidth]{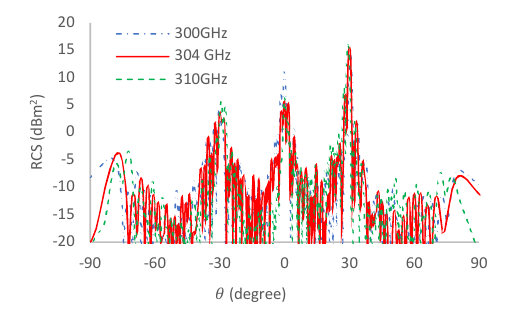}
	\caption{\small{RCS patterns  of the designed $100\times100$ static RIS with $2$-bit phase resolution unit cells.}}
	\label{fig:designed_RIS}
\end{figure}
We have used the Finite Difference Time Domain (FTTD) solver of CST Studio Suite to evaluate the performance of a static RIS comprising $100\times100$ of the previously designed single-layer patch-based unit cells, thus, of aperture $5\times5$ cm$^2$ at $304$ GHz, as shown in Fig. 4. The RCS patterns of this RIS are demonstrated in Fig.~\ref{fig:designed_RIS} for different frequencies. It can observed that the maximum RCS value is $15.6$ dBm$^2$ appearing at the desired non-specular direction of $\theta_{\rm out}=30^\circ$, while it is $6.1$ dBm$^2$ at the specular reflection of $0^\circ$, and $3.9$ dBm$^2$ at the spurious symmetric direction of $-\theta_{\rm out}=-30^\circ$. Clearly, the efficiency of the designed RIS is around $65\%$ when compared to the ideal specular reflection. It is also shown in the figure that our RIS can produce very narrow beams (the Half-Power Beam Width (HPBW) of the main beam is only $1^\circ$), which can be well represented by rays equivalent to plane waves.

\subsection{Far-Field Approximation}
The far-field distance of the designed $5\times5$ cm$^2$ RIS aperture is $10$ m at the studied $304$ GHz band. This implies that, in almost the entire $10\times10\times3$ m$^3$ indoor environment, the RX antenna is in the near field-region of the RIS. Interestingly, our simulations showcased that, after $2$ m, the far-field approximation results in and an amplitude variation below $0.4$ dB. 

\begin{figure}[t!]
	\centering
	\includegraphics[width=0.9\linewidth]{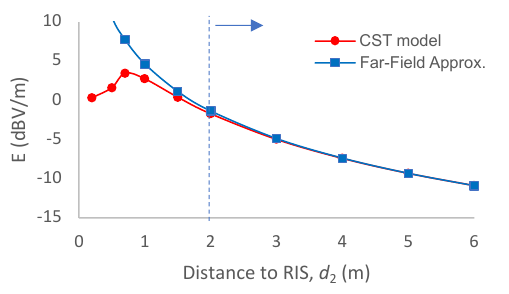}
	\caption{\small{Comparison between $E$-field calculated within CST at different distances from the RIS, considering the far-field approximation (blue curve, square markers) and the more accurate model that includes the near-field (red curve, round markers).}}
	\label{fig:E-field_near_far}
\end{figure}

\section{Indoor Channel Characterization Results}
In this section, we discuss two approximations for the multipath components contributed by the designed static RIS.  

\subsection{Single-Ray Approximation}
\begin{figure}[t!]
	\centering
	\includegraphics[width=0.9\linewidth]{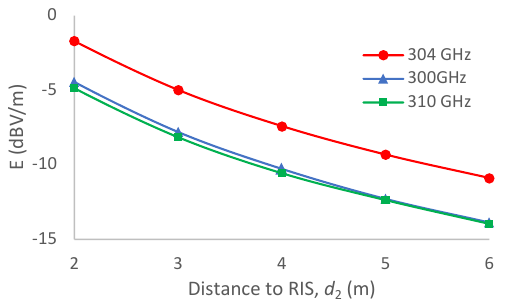}
	\caption{\small{$E$-field versus the RIS-RX distance $d_2$  for a fixed observation angle at $30^\circ$ and different operating frequencies.}}
	\label{fig:E-field}
\end{figure}

\begin{figure}[t!]
    \centering
    \begin{subfigure}[b]{0.9\linewidth}
        \centering
        \includegraphics[width=\linewidth]{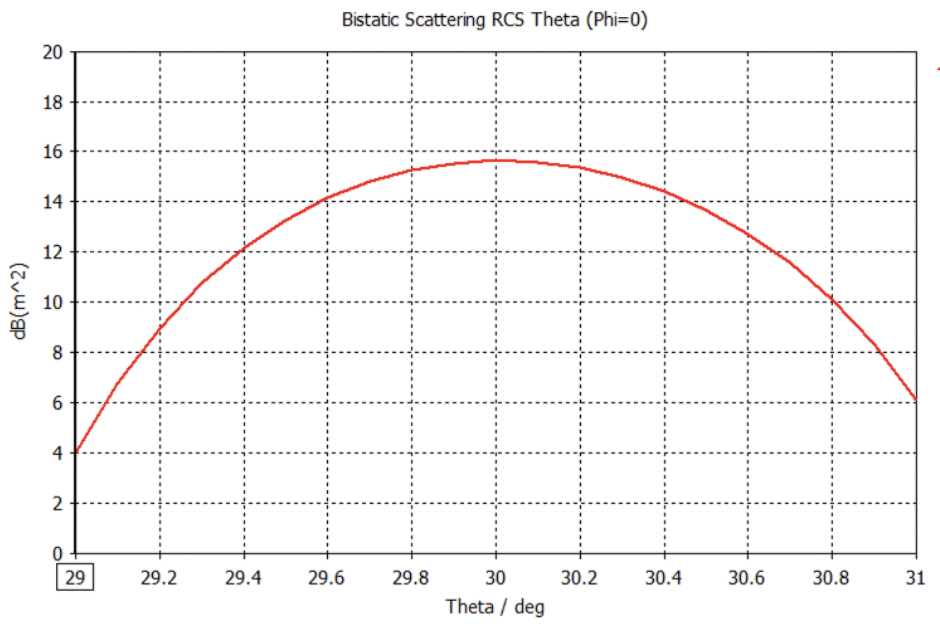}
        \caption{amplitude}
        \label{fig:rcs_amp}
    \end{subfigure}
    
    \bigskip
    
    \begin{subfigure}[b]{0.9\linewidth}
        \centering
        \includegraphics[width=\linewidth]{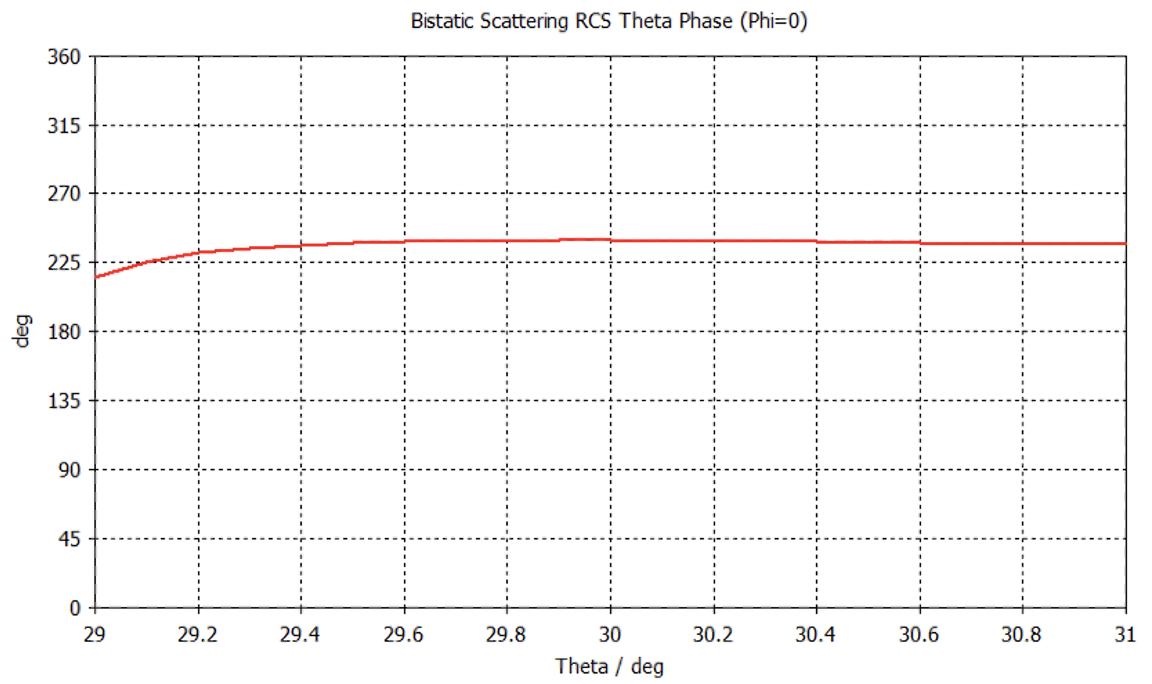}
        \caption{phase}
        \label{fig:rcs_phase}
    \end{subfigure}
    \caption{\small{Full-wave simulations for the RCS amplitude (a) and phase (b) of the designed static $2$-bit RIS.}}
    \label{fig:fullwave_RCS_ris}
\end{figure}

Let us first consider that only a single ray is generated from the RIS. By applying the bistatic radar equation, we can estimate the received power at a given RX position as follows:
\begin{equation}\label{eq:ratio}
\begin{aligned}
    \left( \frac{P_{\rm RX}}{P_{\rm TX}} \right)_{\rm dB} =\, & G_{\rm TX}^{\rm dB} + G_{\rm RX}^{\rm dB} + \sigma_{\rm RIS} \\
    &+ 10\log_{10}\left( \frac{1}{(4\pi)^3}\left( \frac{\lambda}{d_1 d_2} \right)^2 \right),
\end{aligned}    
\end{equation}
where $P_{\rm TX}$ and $P_{\rm RX}$ represent the TX and RX powers, $G_{\rm TX}$ and $G_{\rm RX}$ are the respective antenna gains, $\lambda$ is the wavelength, and $d_1$ and $d_2$ denote the TX-RIS and the RX-RIS distances, respectively, and $\sigma_{\rm RIS}$ is the RCS in dB which has been obtained for the designed static $100\times100$ RIS in Fig.~\ref{fig:designed_RIS}. Note that, in the latter expression, only the amplitude of the RCS is required. It is also important to highlight that the pointing direction of the RIS-generated beam(s) will be affected by the operating frequency according to the following formula~\cite{Longbrake}: 
\begin{equation}\label{eq:Deltatheta}
\Delta\theta(f) = \theta_{\rm out} - \arcsin\left(\frac{f_0}{f}\right),
\end{equation}
where $f_0$ is a nominal frequency. This formula indicates that, for each frequency, a different ray distribution is obtained. It is noted that many of the exciting ray tracing algorithms do not account for this intrinsic dispersive nature of RISs. In Fig.~\ref{fig:E-field}, considering the same observation angle at different RIS-RX distances $d_2$, it is verified that, within a $10$ GHz of bandwidth (i.e., around $3.3\%$ of the considered operating frequencies), the RCS can vary by $3$ dB. It also depicted in Fig.~\ref{fig:fullwave_RCS_ris} that, within the $12$ dB angular range of the main beam direction from the RIS, the phase is nearly constant. This fact validates this single-ray approximation for this region.

\subsection{Multi-Ray Approximation}
\begin{figure}[t!]
	\centering
	\includegraphics[width=0.9\linewidth]{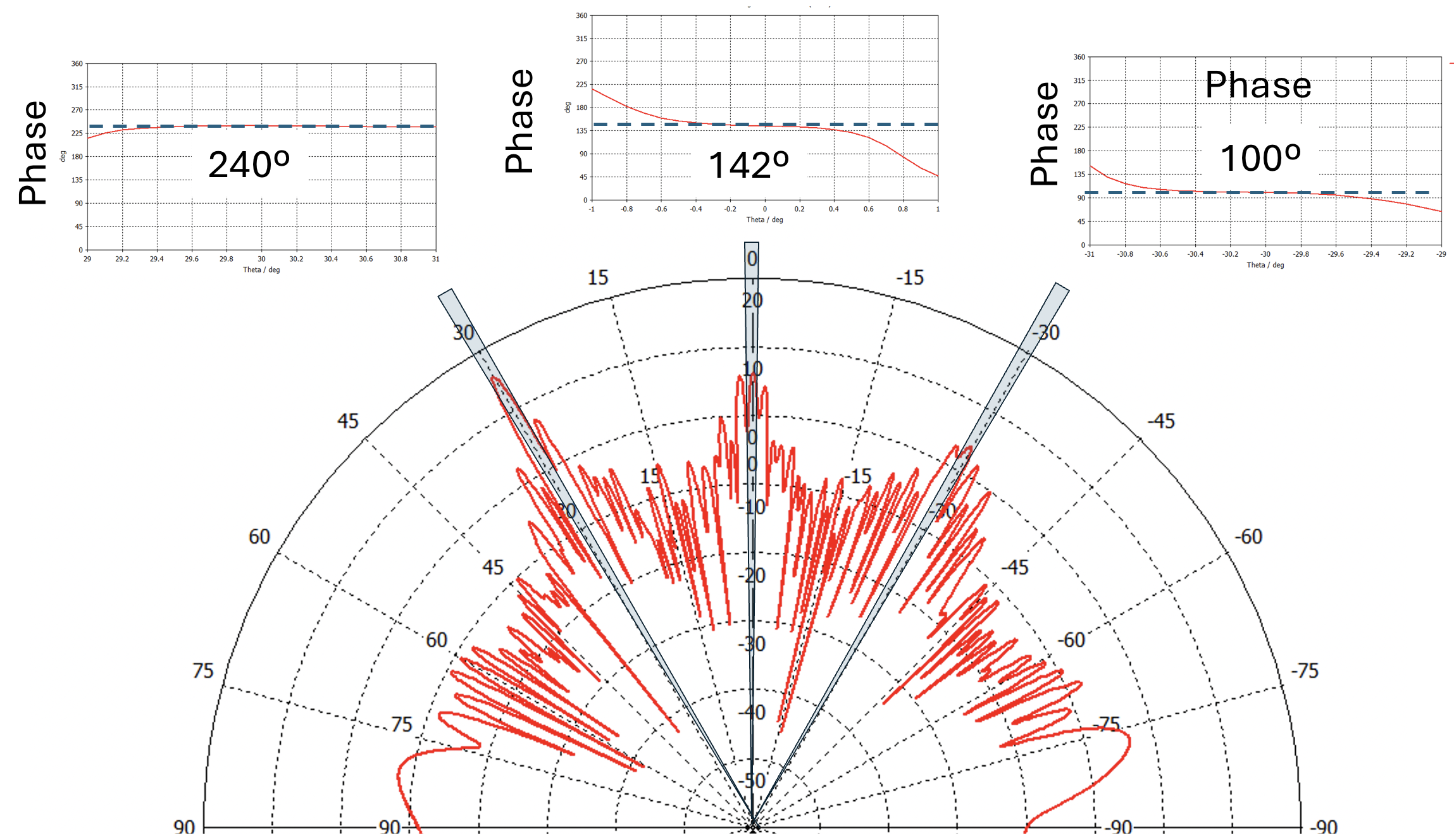}
	\caption{\small{RCS amplitude of the designed static $2$-bit RIS as a function of $\theta$, including the detail of the phase at the main reflection directions.}}
	\label{fig:rcs_cone}
\end{figure}

While it is clear that path-loss will reduce the role of the multipath components at sub-THz frequencies, it is still expected that ray interference will play a significant role on the channel behavior~\cite{Model_THz_data_centers}. In the case of RIS-assisted wireless communications, we have already shown that spurious rays are inevitable parts of the RIS scattering response~\cite{TERRAMETA_Eucap2024}, which will create additional interference paths that need to be accurately accounted for in the channel modeling and characterization. In Fig.~\ref{fig:rcs_cone}, the amplitude and relative phase between all the rays that emerge from the designed RIS when illuminated by a plane wave are demonstrated. It can be seen from the $2^\circ$ cones along the main reflection directions that there exist three dominant rays originating from the RIS. 

\section{Conclusion}
In this paper, we focused on the characterization of an NLOS indoor communication channel at $300$ GHz including our $100\times100$ RIS with $2$-bit resolution single-layer patch-based unit cells, which was designed to provide a static non-specular reflection to incoming normal plane
waves. We have demonstrated that the RIS behavior can be conveniently approximated by a three-ray model, which can be efficiently incorporated within available ray tracing tools. It was also found that the far-field approximation is quite reasonable for much lower RIS-RX distances than those expected in theory.

\section*{Acknowledgment}
This work was supported by the SNS JU project TERRAMETA under the European Union's Horizon Europe research and innovation programme under Grant Agreement No 101097101, including top-up funding by UK Research and Innovation (UKRI) under the UK government’s Horizon Europe funding guarantee.

\bibliographystyle{IEEEtran}
\bibliography{references}
\end{document}